\documentclass{asaproc}

\usepackage{graphicx}
\usepackage{times}

\newcommand{\be}{\begin{equation}}
\newcommand{\ee}{\end{equation}}

\title{Detecting Extra-solar Planets with a Bayesian hybrid MCMC Kepler periodogram}

\author{P. C. Gregory\\
Physics and Astronomy, University of British Columbia\\
e-mail: gregory@phas.ubc.ca\\http://www.physics.ubc.ca/~gregory/gregory.html\\In JSM Proceedings,  Denver, CO, Aug. 3, 2008: American Statistical Association.}

\begin{document}

\maketitle

\begin{abstract}
A Bayesian re-analysis of published radial velocity data sets is providing evidence for additional planetary candidates. The nonlinear model fitting is accomplished with a new hybrid Markov chain Monte Carlo (HMCMC) algorithm which incorporates parallel tempering, simulated annealing and genetic crossover operations. Each of these features facilitate the
detection of a global minimum in $\chi^2$. By combining all three, the HMCMC greatly increases the probability of realizing this goal. When applied to the Kepler problem it acts as a powerful multi-planet Kepler periodogram for both parameter estimation and model selection. The HMCMC algorithm is embedded in a unique two stage adaptive control system that automates the tuning of the MCMC proposal distributions through an annealing operation. 
\end{abstract}

\section{Introduction}

The discovery of multiple planets orbiting the Pulsar PSR B1257+12 (Wolszczan \& Frail, 1992), ushered in an exciting new era of astronomy. Twenty six years later, over 300 extra-solar planets have been discovered by a variety of techniques, including precision radial velocity measurements which have detected the majority of planets to date (Extrasolar Planets Encyclopedia, http://vo.obspm.fr/exoplanetes/encyclo/index.php). From the radial velocity measurements astronomers can detect the reflex motion of a star due to the gravitational tug of its planet(s).
Improvements in precision radial velocity measurements and continued monitoring are permitting the detection of lower amplitude planetary signatures. One example of the fruits of this work is the detection of a super earth in the habital zone surrounding Gliese 581 by Udry et al. (2007). This and other remarkable successes on the part of the observers is motivating a significant effort to improve the statistical tools for analyzing radial velocity data, e.g., Ford \& Gregory (2007), Ford (2006), Ford (2005), Gregory (2005b), Cumming (2004), and Loredo (2004). Much of the recent work has highlighted a Bayesian MCMC approach as a way to better understand parameter uncertainties and degeneracies and to compute model probabilities.

Gregory (2005a \& b; 2007a, b, c) has been developing a Bayesian nonlinear model fitting algorithm based on an MCMC approach that incorporates parallel tempering and simulated annealing to efficiently explore large model parameter spaces. When applied to a Kepler model the prior information insures that any periodic signal detected satisfies Kepler's laws and thus the algorithm functions as a Kepler periodogram~\footnote{Following on from the pioneering work on Bayesian periodograms by Jaynes (1987) and Bretthorst (1988).}. The algorithm when used to search for multiple extra-solar planets simultaneously corresponds to a multi-planet Kepler periodogram. It also yields marginal probability distributions for all the model parameters which provides accurate and quantitative information on parameter uncertainties. Using this approach, Gregory (2007a \& b) reported the detection of a second planet in HD 208487 and two additional planets in HD 11964.
The parallel tempering MCMC algorithm employed in this work includes an innovative two stage adaptive control system that automates the selection of efficient Gaussian parameter proposal distributions through an annealing operation. 

The genetic algorithm provides another approach for searching for a global maximum in the joint probability density distribution, or equivalently minimum $\chi^2$, of a nonlinear model. This paper describes recent progress in incorporating the genetic crossover operation into the MCMC algorithm. The new adaptive hybrid MCMC algorithm (HMCMC) incorporates parallel tempering, simulated annealing and gene crossover operations. The rest of this paper provides an overview of the HMCMC algorithm which is intended for global nonlinear model fitting.

\section{The adaptive hybrid MCMC}

Fig. 1 shows a schematic of the Bayesian nonlinear model fitting program. After specifying the nonlinear model, data and priors, Bayes theorem dictates the target joint probability distribution for the model parameters which can be very complex. To compute the marginals for any subset of the parameters it is necessary to integrate the joint probability distribution over the remaining parameters. In high dimensions, the principle tool for carrying out the integrals is Markov chain Monte Carlo based on the Metropolis algorithm. Our hybrid MCMC algorithm is located at the center of the figure. 

One ingredient of our hybrid MCMC is parallel tempering. Multiple MCMC chains are run in parallel with each chain corresponding to a different temperature. We parameterize the temperature by its reciprocal, $\beta=1/T$ which varies from zero to 1.
The joint probability density distribution for the parameters ($\vec{X}$) of model $M_i$, for a particular chain, is given by
\begin{equation}
p(\vec{X}|D,M_i,I,\beta)= P(\vec{X}|M_i,I)\times p(D|\vec{X}M_i,I)^{\beta}
\label{eq:tempering}
\end{equation}
For parameter estimation purposes 8 chains\\
($\beta=\{0.2,0.3,0.4,0.5,0.6,0.72,0.85,1.0\}$) were employed. At an intervals of 10 iterations, a pair of adjacent chains on the tempering ladder are chosen at random and a proposal made to swap their parameter states. A Monte Carlo acceptance rule determines the probability for the proposed swap to occur (e.g., Gregory 2005a, eq. 12.12). This swap allows for an exchange of information across the population of parallel simulations. In the higher temperature simulations, radically different configurations can arise, whereas in higher $\beta$ (lower temperature) states, a configuration is given the chance to refine itself. The final samples are drawn from the $\beta  = 1$ chain, which corresponds to the desired target probability distribution. For $\beta \ll 1$, the distribution is much flatter. The choice of $\beta$ values can be checked by computing the swap acceptance rate. When they are too far apart the swap rate drops to very low values.

Returning to Fig. 1, the items to the left are the required inputs and the outputs are shown on the right. In this paper, we focus on the parameter estimation aspects of the problem. An important related problem is the calculation of the marginal likelihood required for model selection (e.g., Ford \& Gregory 2007, Gregory 2007b). The bottom of the figure pertains to the adaptive two stage control system (CS) that automatically selects a useful set of Gaussian proposal distributions, one for each parameter. In parallel tempering MCMC, this problem is compounded because of the need for a separate set of Gaussian proposal $\sigma$'s for each chain (different tempering levels). We have automated this process using an innovative two stage statistical control system (CS) in which the error signal is proportional to the difference between the current joint parameter acceptance rate and a target acceptance rate, typically 25\% (Roberts et al. 1997). 

\begin{figure*}
\begin{center}
\includegraphics[width=150mm]{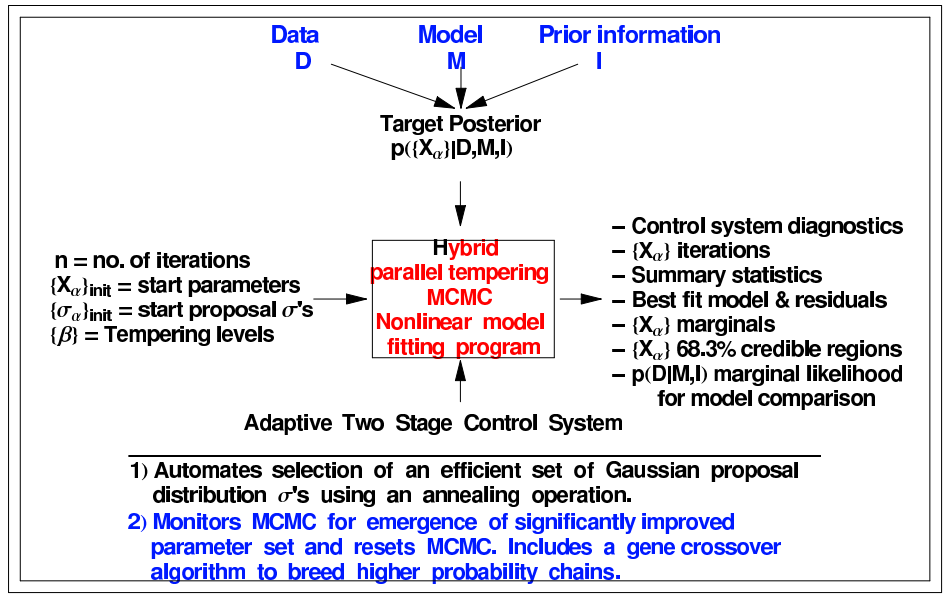}
\caption{Schematic of the Bayesian nonlinear model fitting program which employs the hybrid MCMC algorithm to carry out the Bayesian integrals.}
\end{center}
\end{figure*} 

The first stage CS, which involves annealing the set of Gaussian proposal distribution $\sigma$'s, was described in Gregory 2005a. An initial set of proposal $\sigma$'s ($\approx 10\%$ of the prior range for each parameter) are used for each chain. During the major cycles, the joint acceptance rate is measured based on the current proposal $\sigma$'s and compared to a target acceptance rate. During the minor cycles, each proposal $\sigma$ is separately perturbed to determine an approximate gradient in the acceptance rate for that parameter. The $\sigma$'s are then jointly modified by a small increment in the direction of this gradient. This is done for each of the parallel chains. Proposals to swap parameter values between chains are allowed during major cycles but not within minor cycles. 

The annealing of the proposal $\sigma$'s occurs while the MCMC is homing in on any significant peaks in the target probability distribution. Concurrent with this, another aspect of the annealing operation takes place whenever the Markov chain is started from a location in parameter space that is far from the best fit values. This automatically arises because all the models considered incorporate an extra additive noise whose probability distribution is Gaussian with zero mean and with an unknown standard deviation $s$. When the $\chi^2$ of the fit is very large the 
Bayesian Markov chain automatically inflates $s$ to include anything in the 
data that cannot be accounted for by the model with the current set of 
parameters and the known measurement errors. This results in a smoothing out of the $\chi^2$ surface (Ford 2006) and allows the Markov chain to explore the parameter space more quickly. The chain begins to decrease the value of $s$ as it settles in near the best-fit parameters. An example of this is shown in Fig. 2. In the early stages $s$ is inflated to around 80 m s$^{-1}$ and then decays to a value of $\approx 11$ m s$^{-1}$ over the first 15,000 iterations. This is similar to simulated annealing, but does not require choosing a cooling scheme. 

Although the first stage CS achieves the desired joint acceptance rate, it often happens that a subset of the proposal $\sigma$'s are too small leading to an excessive autocorrelation in the MCMC iterations for these parameters. Part of the second stage CS corrects for this. 
The goal of the second stage is to achieve a set of proposal $\sigma$'s that equalizes the MCMC acceptance rates when new parameter values are proposed separately and achieves the desired acceptance rate when they are proposed jointly. Details of the second stage CS were given in Gregory 2007c and a schematic of the full adaptive control system shown in Fig. 3.
\begin{figure*}
\begin{center}
\includegraphics[width=150mm]{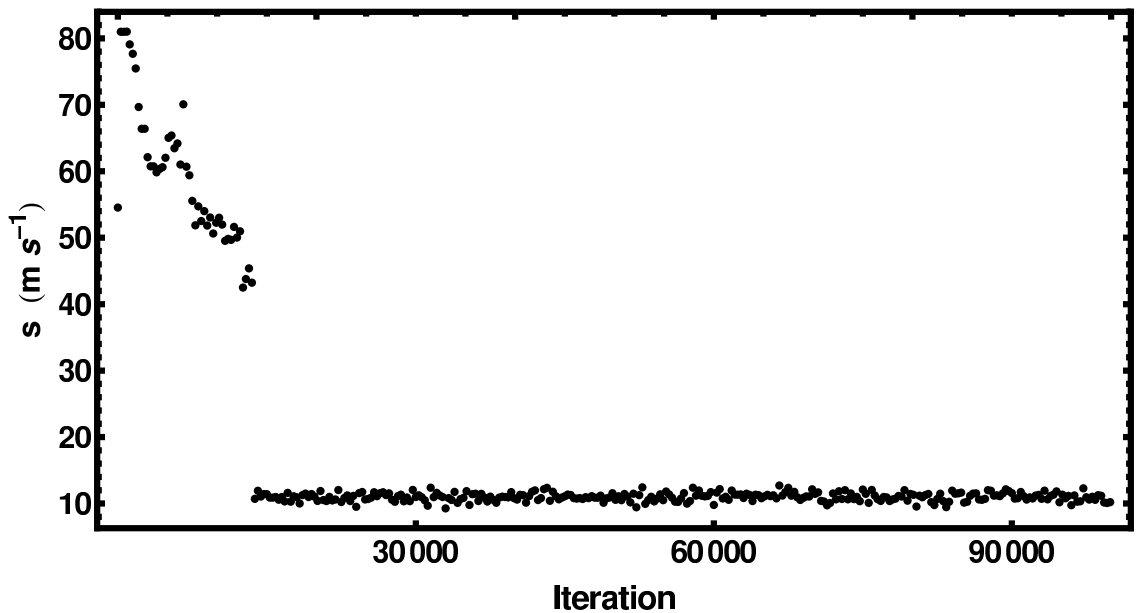}
\caption{The inclusion of an extra noise term $s$ of unknown magnitude gives rise to 
an automatic annealing operation when the Markov chain is started far from the 
best-fit values. Initially $s$ is inflated and then rapidly decays to a much lower level as the best fit parameter values are approached.}
\end{center}
\end{figure*} 

\begin{figure*}
\begin{center}
\includegraphics[width=140mm]{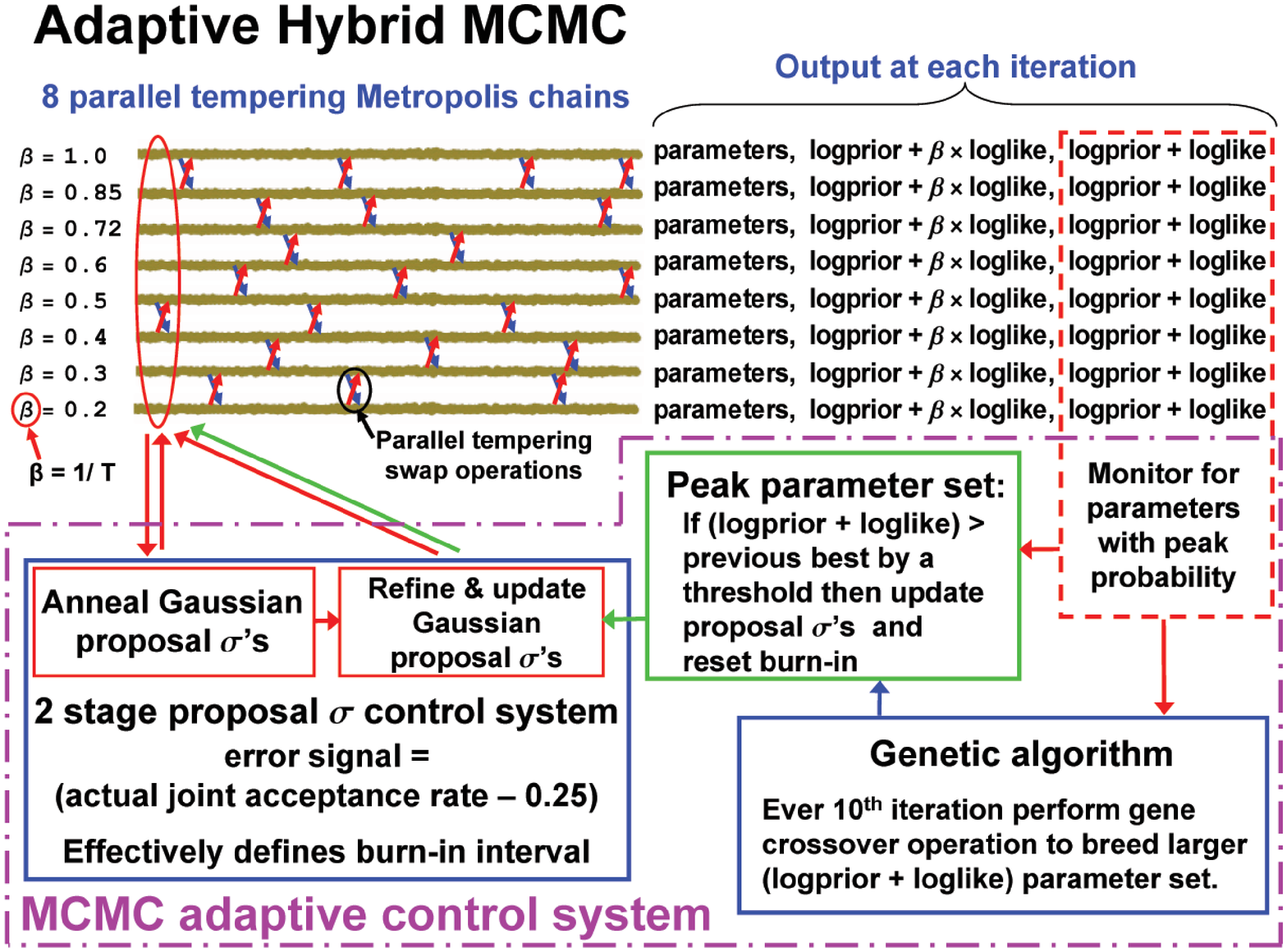}
\caption{Schematic of the operation of the adaptive hybrid MCMC algorithm.}
\end{center}
\end{figure*} 

The first stage is run only once at the beginning, but the second stage can be executed repeatedly, whenever a significantly improved parameter solution emerges. 
Frequently, the burn-in period occurs within the span of the first stage CS, i.e., the significant peaks in the joint parameter probability distribution are found, and the second stage improves the choice of proposal $\sigma$'s based on the highest probability parameter set. 
Occasionally, a new higher (by a user specified threshold) target probability parameter set emerges after the first two stages of the CS are completed. The control system has the ability to detect this and re-activate the second stage. In this sense the CS is adaptive. If this happens the iteration corresponding to the end of the control system is reset.
The useful MCMC simulation data is obtained after the CS are switched off.

The adaptive capability of the control system can be appreciated from an examination of Fig. 3.
The upper left portion of the figure depicts the MCMC iterations from the 8 parallel chains, each corresponding to a different tempering level $\beta$ as indicated on the extreme left. One of the outputs obtained from each chain at every iteration (shown at the far right) is the $\log$ prior $+ \log$ likelihood. This information is continuously fed to the CS which constantly updates the most probable parameter combination regardless of which chain the parameter set occurred in. This is passed to the "Peak parameter set" block of the CS. Its job is to decide if a significantly more probable parameter set has emerged since the last execution of the second stage CS. If so, the second stage CS is re-run which is the basic adaptive feature of the CS.

Recently, a new genetic algorithm block has been added at the bottom right of Fig. 3. The current parameter set can be treated as a set of genes. In the present version, one gene consists of the parameter set that specify one orbit. On this basis, a three planet model has three genes. At any iteration there exist within the CS the most probable parameter set to date $\vec{X}_{\rm max}$, and the most probable parameter set from the most recent iteration $\vec{X}_{\rm cur}$. At regular intervals (user specified) each gene from $\vec{X}_{\rm cur}$ is swapped for the corresponding gene in $\vec{X}_{\rm max}$. If either substitution leads to a higher probability it is retained and $\vec{X}_{\rm max}$ updated. The effectiveness of this operation can be tested by comparing the number of times the gene crossover operation gives rise to a new value of $\vec{X}_{\rm max}$ compared to the number of new $\vec{X}_{\rm max}$ arising from the normal parallel tempering MCMC iterations. The gene crossover operations prove to be very effective, and give rise to new $\vec{X}_{\rm max}$ values $\approx 3$ times more often than MCMC operations. Of course, most of these swaps lead to very minor changes in probability but occasionally big jumps are created. Not surprisingly, it turns out that individual gene swaps from $\vec{X}_{\rm cur}$ to $\vec{X}_{\rm max}$ are much more effective than the other way around (reverse swaps) by a factor of about 17. Since it costs just as much time to compute the probability for a swap either way we no longer carry out the reverse swaps. Instead, we have extended this operation to swaps from $\vec{X}_{\rm cur2}$, the parameters of the second most probable current chain, to $\vec{X}_{\rm max}$. This gives rise to new values of $\vec{X}_{\rm max}$ at a rate approximately half that of swaps from $\vec{X}_{\rm cur}$ to $\vec{X}_{\rm max}$. Further experimentation with this concept is ongoing, in particular concerning which scheme produces the largest average change in probability per iteration.    

\section{Results}

The addition of the genetic block has definitely increased the likelihood of detecting a global maximum in the joint parameter probability distribution. The algorithm has been very successful in carrying out blind searches for a three planet model in typical sparce exoplanet data sets and some success has been obtained with 4 planet searches as well. Fig. 4 shows the results of a blind three planet search for two different trials on HD 11964 (87 radial velocity measurements) which we previously demonstrated has 3 planets (Gregory 2007b) with periods of 38, 360, and 1924 d. The different starting set of periods are given in the figure caption. 
\begin{figure*}
\begin{center}
\begin{tabular}{c@{\qquad}c}
\mbox{\includegraphics[width=85mm]{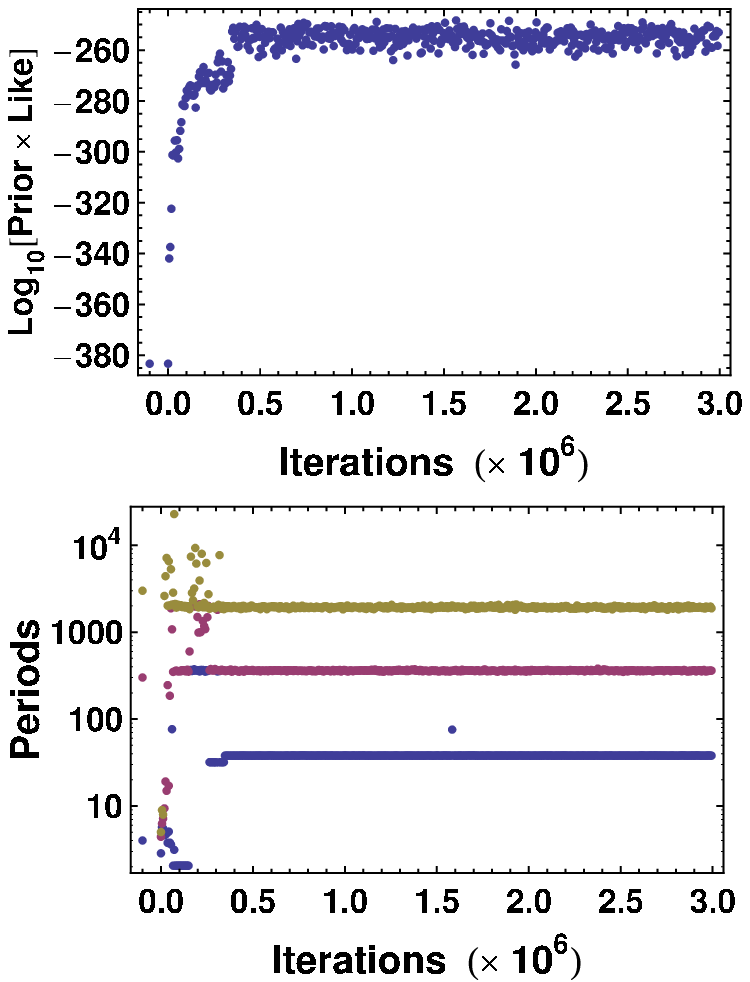}}&
\mbox{\includegraphics[width=85mm]{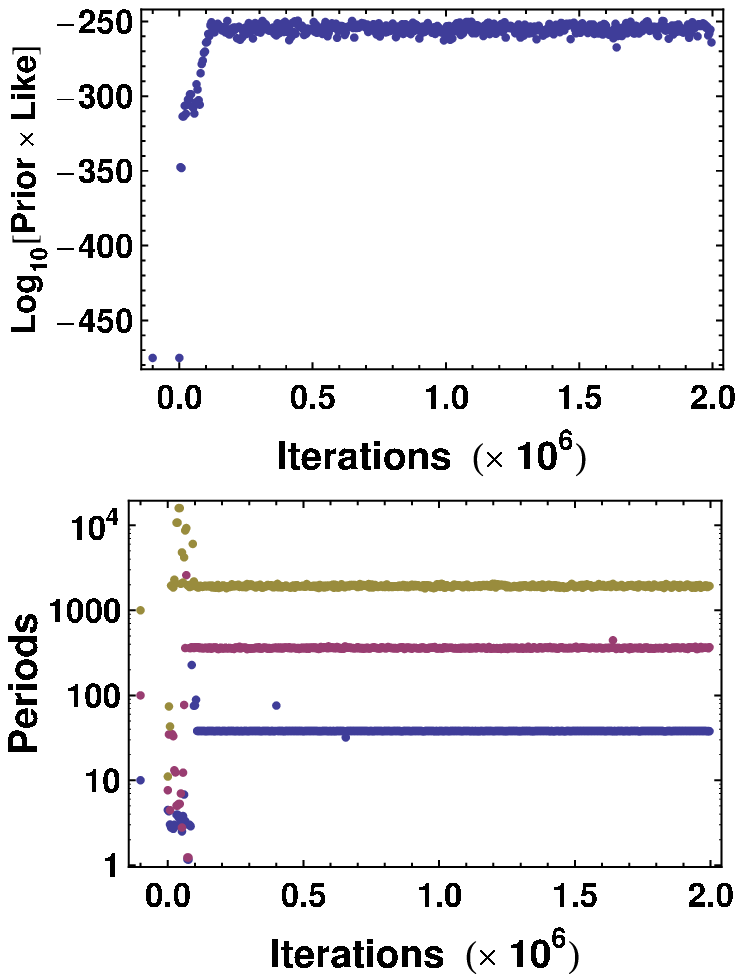}}
\end{tabular}
\end{center}
\caption{Three planet blind search for star HD 11964 for two different trials. In each case the upper panel shows the evolution of the Log$_{10}[$prior$\times$likelihood$]$ versus MCMC iteration and the lower panel the evolution of the three periods found. The three starting periods were 4, 300, 3000 d for the left hand trial and 10, 100, 1000 d for the right.}
\end{figure*}

Recently, we experimented with a blind 5 planet search in a sample of 300 measurements for 55 Cancri, kindly provided by Debra Fischer. This system is known from there work to contain 5 planets with periods of 2.8, 14.6, 44.4, 260, and 5228 d (Fischer et al. 2008). The left hand panel of Fig. 5 shows the 5 periods detected during a 5 planet blind search of the 55 Cancri data, starting from initial periods of 2, 10, 100, 300, 3000 d. The program found 4 of the 5 previously published periods. 
The right hand panel shows the periods found in a 5 planet blind search when the period of the strongest component at 14.65d was specified and an upper limit imposed on all the planetary eccentricities of 0.4. The complete set of initial periods values was 2, 14.65, 100, 500, 5000 d. In this case the program found all 5 of the previously published periods.
\begin{figure*}
\begin{center}
\begin{tabular}{c@{\qquad}c}
\mbox{\includegraphics[width=85mm]{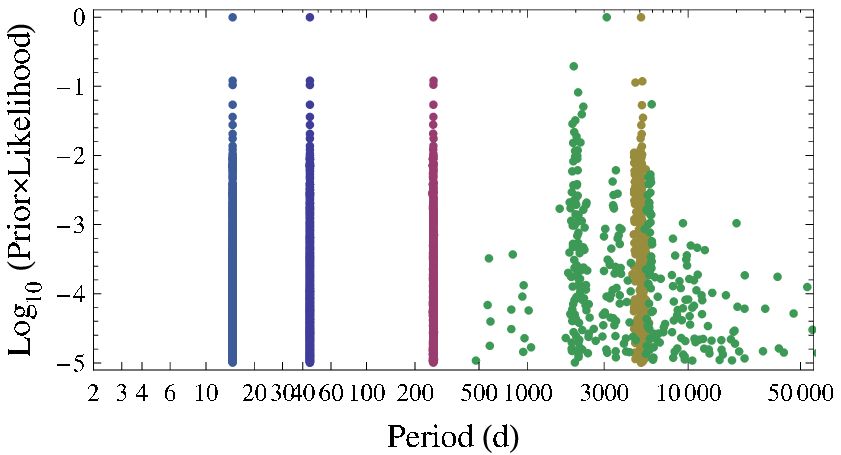}}&
\mbox{\includegraphics[width=85mm]{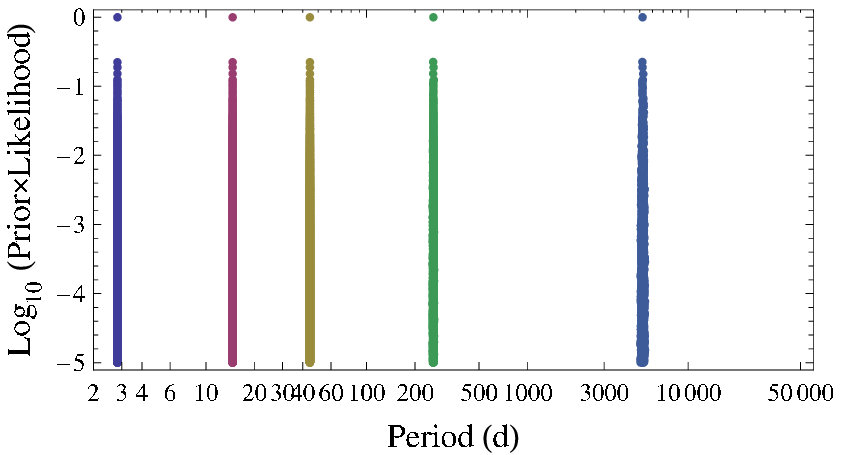}}
\end{tabular}
\end{center}
\caption{The left hand plot shows the periods found in a 5 planet blind search of a sample of 55 Cancri data from initial periods of 2, 10, 100, 300, 3000 d. The program found 4 of the 5 previously published periods.
The right hand plot shows the periods found in a 5 planet blind search when the period of the strongest component at 14.65 d was specified and an upper limit on all planetary eccentricities set at 0.4. The complete set of initial periods values was 2, 14.65, 100, 500, 5000 d. In this case the program found all 5 of the previously published periods.}
\end{figure*}

\begin{figure*}
\centering
\includegraphics[scale=0.5]{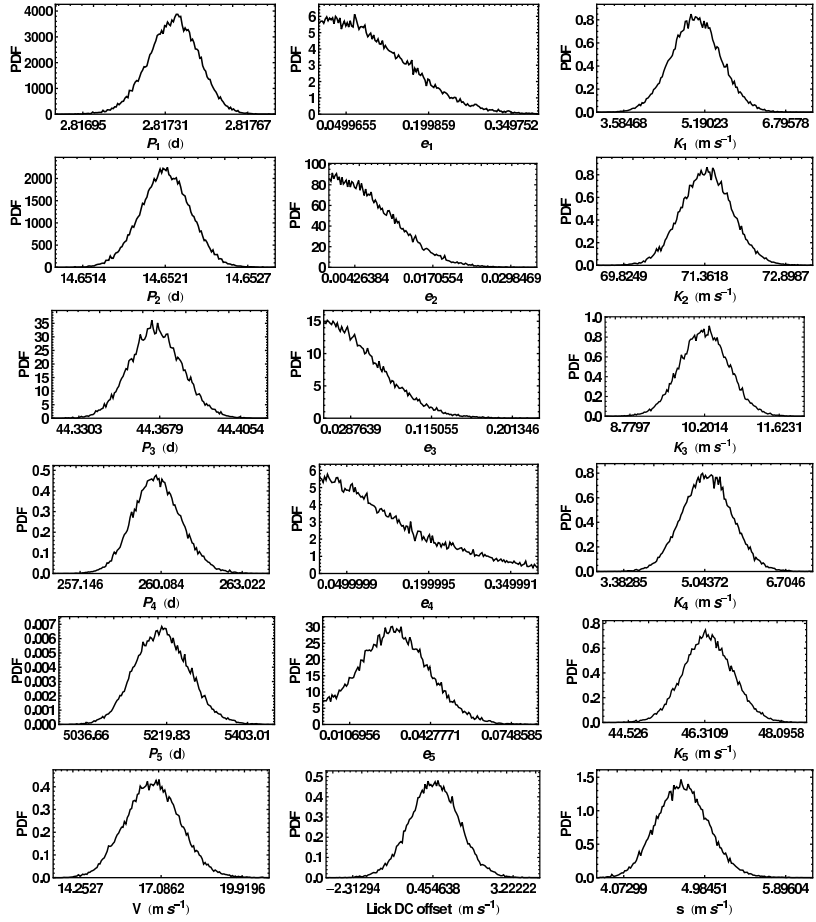}
\caption{Five planet marginal probability distributions for a selection of parameters from the 55 Canri analysis.}
\end{figure*}

\begin{figure*}
\centering
\includegraphics[scale=0.5]{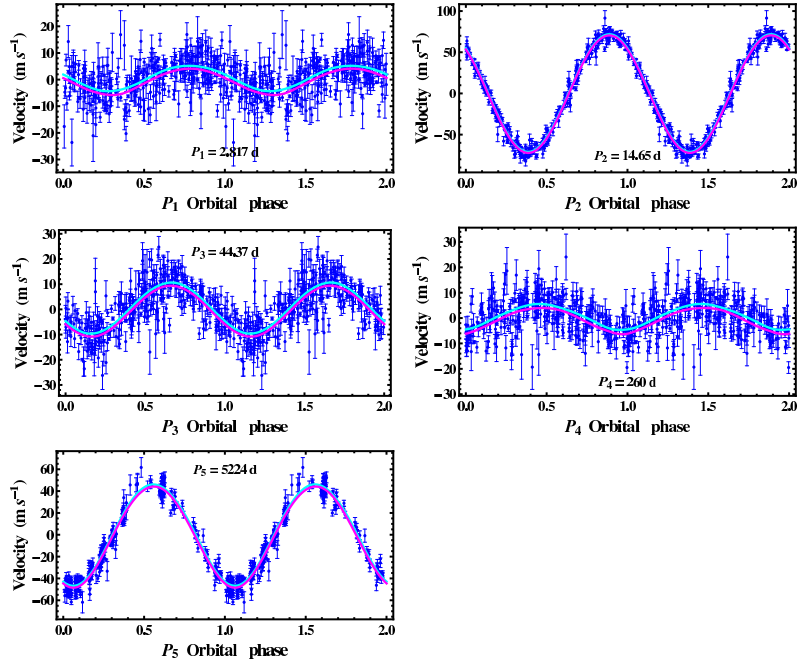}
\caption{55 Cancri phase plots. In each panel, the data with the mean model of four of the five orbits subtracted, is plotted versus phase computed from the MAP value of the remaining period.  Overlaid on this is the mean RV curve $+1$ standard deviation (cyan) and the mean $-1$ standard deviation (magenta), computed from the MCMC parameter iterations for the particular orbit.}
\end{figure*}

Fig. 6 shows the marginal probability distributions for a selection of the 5 planet model parameters which include the period, eccentricity, and amplitude parameter $K$. The bottom three panels are $V$, the systematic velocity of the star, the unknown velocity offset of the Lick spectrometer relative the Keck spectrometer, and $s$ the extra noise parameter. The peak of $s = 4.9$ m s$^{-1}$ hints at the possibility of one or more additional planets.

Fig. 7 shows 55 Cancri phase plots. The 5 planet parameter set corresponding to any single post burn-in MCMC iteration can be used to construct a model radial velocity curve. We can compute the mean and standard deviation of these model curves. In each panel, the data with the mean model of four of the five orbits subtracted, is plotted versus phase computed from the MAP value of the remaining $5^{\rm th}$ period.  Overlaid on this is the mean RV curve $+1$ standard deviation (cyan) and the mean $-1$ standard deviation (magenta), computed from the MCMC parameter iterations for the $5^{\rm th}$ orbit.

\section{Concluding remarks}

In this paper we have focused on describing the operation of the adaptive hybrid MCMC. It integrates the advantages of parallel tempering, simulated annealing and the genetic algorithm. Each of these techniques was designed to facilitate the detection of a global minimum in $\chi^2$. Combining all three in an adaptive hybrid MCMC greatly increase the probability of 
realizing this goal. Many of the details pertinent to exoplanet analysis concerning the model equations, the mathematical form of the target distribution (Gregory 2005b), parameter priors and a useful re-parameterization (Ford 2006, Gregory 2007a, 2007b), have been left out. In particular, a multi-frequency prior suitable for simultaneous multiple planet searches is derived in Gregory 2007a together with a discussion of an efficient search strategy. 

The adaptive Bayesian hybrid MCMC is very general and can be applied to many different nonlinear modeling problems. It has been implemented in {\it gridMathematica} on an 8 core PC. The increase in a speed for the parallel implementation is a factor 6.6. When applied to the Kepler problem it corresponds to a multi-planet Kepler periodogram which yields full marginal distributions for all the orbital parameters. The execution time for a 1 planet blind fit (7 parameters) is $10^6$ iterations per hr. The program scales linearly with the number of parameters being fit (28 parameters for the 5 planet 55 Cancri fit). As discussed in Gregory 2007b the Kepler periodogram employs an alternative method for converting the time of an observation to true anomaly that enables it to handle much larger data sets without a significant increase in computation time. For example, an increase in the data by a factor of 6.5 resulted in only an 18\% increase in execution time. 

The author would like to thank Wolfram Research for providing a complementary license to run {\it gridMathematica} that was used in this research.

\end{document}